\documentstyle[prl,aps,preprint]{revtex}

\begin{document}

\draft
\title{Unpolarized quasielectrons and the spin polarization at 
filling fractions between $\nu=1/3$ and $\nu=2/5$}

\author{Jacek Dziarmaga\thanks{E-mail: {\tt J.P.Dziarmaga@durham.ac.uk}}
and Meik Hellmund\thanks{E-mail: 
{\tt Meik.Hellmund@itp.uni-leipzig.de}}\thanks{
permanent address: Institut
f{\"u}r Theoretische Physik, Leipzig, Germany}}

\address{Department of Mathematical Sciences,
         University of Durham, South Road, Durham, DH1 3LE,
         United Kingdom}
\date{April 30, 1997}
\maketitle
\tighten

\begin{abstract}

 We prove that for a hard core interaction the ground state spin
polarization in the low Zeeman energy limit is given by $P=2/\nu-5$ for
filling fractions in the range $ 1/3 \leq\nu\leq 2/5 $. The same result
holds for a Coulomb potential except for marginally small magnetic
fields.  At the magnetic fields $B<20T$ unpolarized quasielectrons can
manifest themselves by a characteristic peak in the I-V characteristics
for tunneling between two $\nu=1/3$ ferromagnets. 

\end{abstract}

\section{Introduction}
 
Recent results\cite{pol} on spin depolarization at quantum Hall systems in
the vicinity of filling fraction $1$ provoke questions about the spin
polarization at other filling fractions.  A potentially interesting range
of filling fractions is the interval $8/5\leq\nu\leq 5/3$, which is
equivalent to $1/3\leq\nu\leq2/5$ by particle-hole symmetry $\nu
\leftrightarrow 2-\nu$.  The $\nu=1/3$ state is a ferromagnetic Laughlin
state\cite{L}, while the $\nu=2/5$ is a ferromagnet or a spin-singlet
depending on the Zeeman energy. For low Zeeman energy the question arises
how the spin polarization changes between the limiting ferromagnetic and
spin-singlet states. In this paper we give an answer to this question. 

\section{Spin polarization in the range $1/3<\nu\leq 2/5$ for a 
          hard core potential }

   The ferromagnetic $\nu=1/3$ ground state is described by the celebrated
Laughlin wave function
 
   \begin{equation}\label{10}
   \psi_{1/3}(z_{k})=
   \prod_{m>n=1}^{N}(z_{m}-z_{n})^{3} \;
   |\downarrow_{1}\ldots\downarrow_{N}> \;\;.
   \end{equation}
This state is an exact zero energy eigenstate of the hard core potential
~\cite{TK}

   \begin{equation}\label{20}
   V(z)=\infty \delta(z)+\lambda\nabla^{2}\delta(z)
   \end{equation}
with Haldane's\cite{Hald83} pseudopotentials $V_{0}=\infty,V_{1}>0$ and
$V_{k}=0$ for $k\geq 2$. It is the unique zero energy eigenstate for
arbitrarily small but nonzero Zeeman energy. 

  For vanishing Zeeman energy the unique zero energy eigenstate at the
filling fraction of $2/5$ is~\cite{Rezayi} the Halperin's spin-singlet
$(3,3,2)$ state~\cite{Halperin}
 
   \begin{equation}\label{30}
   \psi_{2/5}(z_{k})=
   \prod_{i>j=1}^{N/2}(z_{i}-z_{j})^{3}
   \prod_{k>l=N/2}^{N}(z_{k}-z_{l})^{3}
   \prod_{m=0}^{N/2} \prod_{n=N/2}^{N} (z_{m}-z_{n})^{2} \;
   |\uparrow_{1}\ldots\uparrow_{N/2}\downarrow_{N/2+1}\ldots\downarrow_{N}> 
   \;\;.
   \end{equation}
We display only one spinor component, the one where the spins of the
electrons $1,..,N/2$ are pointing up and those of the electrons
$N/2+1,\ldots,N$ are pointing down. 

  $\nu=1/3$ is the highest filling fraction for which the polarized
ground state can still have zero energy\cite{TK,Rezayi}. For $\nu >1/3$
electrons are packed together too close, such that some pairs of
electrons have to be in a state of relative angular momentum $1$. This
effect gives rise to the discontinuity of the chemical potential at
$\nu=1/3$. More space can be created by reversing some number of spins. 
For one flux quantum less than at $\nu=1/3$, the lowest angular momentum
zero energy eigenstate is the unpolarized quasielectron~\cite{Rezayi}
with one reversed spin
 
  \begin{equation}\label{40}
  \prod_{k=2}^{N}
  (z_{k}-z_{1})^{2}
  \prod_{m>n=2}^{N} (z_{m}-z_{n})^{3} \;
  |\uparrow_{1}\downarrow_{2}\ldots\downarrow_{N}> \;\;.
  \end{equation}
The relative angular momenta are at least $3$ for pairs of electrons of
the same spin and at least $2$ for those of opposite spin. The total spin
of the state (\ref{40}) is $S=\frac{N}{2}-1$. All zero energy states of
this total spin can be obtained by multiplication of the wave function
(\ref{40}) with a polynomial which is symmetric in the coordinates of the
spin-up electrons.  In this case the polynomials are factors $z_{1}^{k}$
with $k=0,\ldots,N-2$. Further zero energy states can be constructed with
more reversed spins, $S<\frac{N}{2}-1$. 
 
  For two flux quanta less than at $\nu=1/3$, a polarized ground state
contains two polarized quasielectrons. In order to have zero energy,
some spins must be flipped. Just one flipped spin is not enough, even if
the relative angular momenta of the same spin electrons are at least
$3$, as in the state
 
  \begin{equation}\label{50}
  \prod_{k=2}^{N}
  (z_{k}-z_{1})
  \prod_{m>n=2}^{N} (z_{m}-z_{n})^{3} \;
  |\uparrow_{1}\downarrow_{2}\ldots\downarrow_{N}> \;\;,  
  \end{equation}
where pairs of opposite spin still can have the relative angular
momentum of $1$. Therefore the number of reversed spins $R$ has to be at
least $R=2$. The zero energy eigenstate of lowest angular momentum with
$S=\frac{N}{2}-2$ is
 
  \begin{equation}\label{60}
  (z_{1}-z_{2})^{3}
  \prod_{k=3}^{N}
  (z_{k}-z_{1})^{2} (z_{k}-z_{2})^{2}
  \prod_{m>n=3}^{N} (z_{m}-z_{n})^{3} \;
  |\uparrow_{1}\uparrow_{2}\downarrow_{3}\ldots\downarrow_{N}> \;\;.
  \end{equation}
All other zero energy states of the same total spin are obtained by
multiplication of Eq.(\ref{60}) with polynomials symmetric in
$z_{1},z_{2}$, $P_{a,b}=(z_{1}+z_{2})^{a}(z_{1}-z_{2})^{2b}$, such that
$a+2b \leq N-1$. There exist further zero energy eigenstates with more
reversed spins, $S<\frac{N}{2}-2$. 

  In general, for $\Phi$ flux quanta less than at $\nu=1/3$, at least
$\Phi$ spins have to be reversed in order to get a zero energy eigenstate.
All zero energy eigenstates of the spin $S=\frac{N}{2}-\Phi$ are obtained
from a minimal angular momentum seed state
 
  \begin{equation}\label{70}
  \prod_{i>j=1}^{\Phi} (z_{i}-z_{j})^{3}
  \prod_{k=\Phi+1}^{N}
  \prod_{l=1}^{\Phi}
  (z_{k}-z_{l})^{2}
  \prod_{m>n=\Phi+1}^{N} (z_{m}-z_{n})^{3} \;
  |\uparrow_{1}\ldots\uparrow_{\Phi}
   \downarrow_{\Phi+1}\ldots\downarrow_{N}> \;
  \end{equation}
by multiplication with symmetric polynomials of $z_{1},\ldots,z_{\Phi}$. 
Further zero energy eigenstates can be constructed for
$S<\frac{N}{2}-\Phi$. In the limit of $\nu=2/5$ or $\Phi=\frac{N}{2}$, the
spin singlet (\ref{30}) is the unique zero energy eigenstate. 

  To summarize, for any filling fraction in the range from $1/3$ to
$2/5$, which we characterize by the number $\Phi$ of flux quanta
relative to $\nu=1/3$, there is a degenerate band of zero energy
eigenstates with spins $S=0,\ldots,\frac{N}{2}-\Phi$. The states with $S
> \frac{N}{2}-\Phi$ have a nonzero interaction (\ref{20}) energy. The
degeneracy of the zero energy band is partially removed by arbitrarily
small but nonzero Zeeman energy. Then the lowest states are those with
$S=\frac{N}{2}-\Phi$. For nonzero Zeeman energy the spin of the ground
state is not lower than $S=\frac{N}{2}-\Phi$. Thus for the hard core
model (\ref{20}) the polarization $P\equiv\frac{2S}{N}$ of the ground
state at filling fraction $\nu$ is bounded from below by

  \begin{equation}\label{80}
  P\geq \frac{2}{\nu}-5 \;\;.
  \end{equation}
A similar bound holds, by particle-hole symmetry, for $\nu\in (8/5,5/3)$. 
The bound is saturated for small Zeeman energy. Eq.(\ref{80}) is a
rigorous result for the hard core potential. In the following section we
discuss its relevance for realistic Coulomb interaction.

\section{ Spin polarization in the range $1/3<\nu\leq 2/5$ for
          Coulomb potential}

  The hard core potential (\ref{20}) captures important characteristics of
the Coulomb potential. For instance, the states (\ref{10}) and (\ref{30})
are almost identical with Coulomb ground states (overlaps better than
0.99). The Coulomb potential is long ranged, however, so that its higher
Haldane pseudopotentials are nonzero albeit small. This tail shifts all
the energy levels and removes the degeneracy of the hard core potential's
zero energy band. As we showed in the preceding section, a nonzero
pseudopotential $V_{1}$ forced a minimal number of spins to be reversed.
For a Coulomb potential this tendency goes even further - the more spins
are reversed the better. For $\Phi$ flux quanta less than at $\nu=1/3$,
the former band of zero energy with $S=0,\ldots,\frac{N}{2}-\Phi$ is no
longer degenerate. Instead, the states with lower spin $S$ have lower
Coulomb energy. As the higher quasipotentials are weak we can expect this
splitting to be small. 

  Let us assume that the splitting of the hard core zero energy band is
small indeed. In the strong Zeeman energy limit the ground state is
polarized for any filling fraction. As the Zeeman energy is decreased the
spin of the ground state decreases relatively quickly from $S=\frac{N}{2}$
to $S=\frac{N}{2}-\Phi$. Within this range the scale for the gain of Coulomb
energy per flipped spin is set by $V_{1}-V_{2}$ and further gains of
Coulomb energy are much smaller. If we take into account that Coulomb
energy depends on the magnetic field like $\sqrt{B}$, further depolarization
may require unrealistically small magnetic fields.  We discuss the width
of this split of the zero energy band of the hard core potential in 
two limiting cases. 

\subsection{ Filling fraction just below $\nu=2/5$ }

  In this region the relevant degrees of freedom are quasihole excitations
of the state (\ref{30}) 

  \begin{equation}\label{200}
  \prod_{k=1}^{N/2} (z_{k}-w_{\uparrow}) \psi_{2/5}(z_{k}) \;,
  \;\; \prod_{k=(N/2)+1}^{N} (z_{k}-w_{\downarrow}) \psi_{2/5}(z_{k}) 
  \end{equation}
in the spin-up and spin-down fluid respectively. $N$ is assumed to be
large. For one flux quantum more than at $\nu=2/5$, $\Phi=\frac{N}{2}-1$,
two such quasiholes have to be created.  Let us study this case in more
detail. 

  Creation of a quasihole costs some exchange energy, which does not
depend on the quasihole's spin.  If we set some definite large distance
between the quasiholes, their Coulomb energy will not depend on whether
they are of the same or of opposite spin. Thus the Coulomb interaction
does not favor any spin polarization. It is the Zeeman energy that favors
states with both quasiholes in the spin-up fluid so that the ground
state's spin is $S=1$. However, if we force the quasiholes to be closer
than the magnetic length, the exchange energy will favor a pair of
quasiholes in different fluids so that $S=0$ for the Coulomb ground state. 
 
  The same logic can be applied to a finite density of quasiholes. If the
average distance between quasiholes is larger than the magnetic length,
states with different spin have the same Coulomb energy.  Splitting may
become significant only at a higher density of holes. To get an idea of
how significant it is, we studied the case of one flux quantum less than
at $\nu=1/3$. 

\subsection{ Filling fraction just above $\nu=1/3$ }

  The unpolarized quasielectron (\ref{40}) with one reversed spin,
$S=\frac{N}{2}-1$, has lower Coulomb energy than its polarized
counterpart, $S=\frac{N}{2}$. In order to estimate the energy gain for
this depolarization we performed a finite size study of electrons in
spherical geometry\cite{Hald83}. The one quasiparticle sector at
$\nu=1/3$ is characterized by $3(N-1)-1$ flux quanta piercing the
sphere. We have calculated the lowest eigenstates for Coulomb
interaction in the subspace of states with $S_z=\frac{N}{2}-1$ for up to
9 electrons. The lowest energy state has always $S=N/2-1$, the next
state $S=N/2$. The energy differences are given in the second column of
the Table~\ref{t1}. The data allow fits with quadratic polynomials in
$1/N$ giving the energy gain for $N\to\infty$ of $0.0278(5)
e^2/\varepsilon l$. Therefore, the unpolarized $R=1$ quasielectron is
found to be stable up to remarkably high magnetic fields $B<22T$. Here
we used parameters appropriate to GaAs ( $g \mu_B B \approx 0.3 K \times
B[T], e^2/\epsilon l \approx 50K \times \sqrt{B[T]}$ ). These results
are in agreement with the results in \cite{morf}. Similar calculations
have been performed in the subspace $S_z=N/2-2$ for up to 8 electrons.
The energy gains for the second depolarization are listed in the third
column of the Table~\ref{t1}. The $R=2$ skyrmion becomes stable for
$B<0.75 T$, which is remarkably small as compared to the critical
magnetic field for the first depolarization.

\subsection{Exact diagonalizations for $N=8$ electrons}

  To substantiate the discussion above we have performed exact
diagonalizations for $N=8$ electrons with Coulomb interaction in spherical
geometry.  This gives an estimate for the splitting pattern of the zero
energy band of the hard core model. Variation of the Zeeman energy induces
a spectral flow. At very high Zeeman energy the ground state is fully
polarized ($S=4$). Lowering the Zeeman energy, it gives place to
depolarized states with lower Coulomb energy.  Table~\ref{t2} lists the
magnetic fields at which the states in the hard core zero energy band
become stable. For each filling fraction (or $\Phi$) the highest spin
state $S=N/2-\Phi$ in the band is achieved at remarkably high magnetic
field but further depolarizations require much lower magnetic fields,
which can not be achieved in realistic samples. This property reflects the
fact that for the Coulomb interaction the pseudopotential difference
$(V_{1}-V_{2})$ is large as compared to further differences
$(V_{2}-V_{3}),(V_{3}-V_{4})$, which makes the Coulomb potential similar
to the model hard-core interaction.

\section{ Unpolarized quasielectron's signature in tunneling }

  Tunneling signature of the $R=1$ skyrmion at $\nu=1$ has been studied
in the recent paper ~\cite{tunnel}. The $R=1$ skyrmion can be
interpreted as a bound state of three elementary objects: two holes in
the $\nu=1$ ferromagnet and one spin-up electron. Tunneling of either
spin-up or spin-down electrons leads to characteristic features in the
I-V characteristics. 
 
  The $R=1$ unpolarized quasielectron (UQ1) can also be interpreted as a
bound state of three objects: two quasiholes of charge $+e/3$ in the
$\nu=1/3$ ferromagnet and one spin-up electron of charge $-e$. This
interpretation becomes clear, if we compare the UQ1 wave function
(\ref{40}) with the wave function for a Laughlin quasihole localized at
$w$

\begin{equation}\label{lqh}
\prod_{k=2}^{N} (z_{k}-w) \prod_{m>n=2}^{N} (z_{m}-z_{n})^{3}\;
|\downarrow_{2}\ldots\downarrow_{N}> \;\;.
\end{equation}

  In the wave function (\ref{40}) there are two Laughlin quasiholes,
both of them follow the spin-up electron, $w_{1}=w_{2}=z_{1}$. The
quasiholes are in the state of relative angular momentum $L_z=-1/3$,
described by the pseudo-wavefunction $(\bar{w}_{1}-\bar{w}_{2})^{1/3}$. 

  The spin-up electron constituent of the UQ1 can tunnel to another
isolated $\nu=1/3$ ferromagnet. The final state is a state with a
$L_z=-1/3$ pair of quasiholes on one side of the barrier and the spin-up
electron added to the $\nu=1/3$ spin-down ferromagnet on the other side.
The energy increases by the difference between the gross energy of two
quasiholes and the gross energy of the UQ1. The minimal voltage $V$
required for such a tunneling to take place is given by
$eV=[2\varepsilon_{-}^{(1/3)}-\varepsilon_{UQ1}+E_{-1/3}]\frac{e^{2}}{\kappa
l}$. $\varepsilon_{-}^{(1/3)}\approx 0.231 \frac{e^{2}}{\kappa l}$
according to \cite{mh}. The gross energy of the polarized quasielectron
can be estimated from above~\cite{mh} by $\varepsilon_{+}^{(1/3)}\leq
-0.128 \frac{e^{2}}{\kappa l}$, what, together with our estimate of the
depolarization energy, gives $\varepsilon_{UQ1}\leq -0.156
\frac{e^{2}}{\kappa l}$. $E_{-1/3}$ is the interaction energy of the two
quasiholes in the $L_z=-1/3$ state, which can be estimated to be
$0.040\frac{e^{2}}{\kappa l}$. Therefore the tunneling of the spin-up
electron occurs at the voltage $V\approx 0.66 \frac{e}{\kappa l}$. 

\section{ Conclusions }

  The $R=1$ quasielectrons above $\nu=1/3$ (or below $\nu=5/3$) become
stable for $B<20T$. Further depolarizations, $R>1$, would require
magnetic fields less than $0.75T$. These $R=1$ quasielectrons dominate
the depolarization pattern above $\nu=1/3$. For low Zeeman energy the
number of reversed spins is proportional to the number of
quasielectrons. 

  We expect the following to happen in a sample as in Ref.~\cite{exp}.
At $\nu=8/5$ and $B=5T$ the ground state is a spin singlet so that
$P=0$. A decrease of the magnetic field from $B=5T$ to $B=4.8T$
increases the filling fraction from $\nu=8/5$ to $\nu=5/3$. At the same
time the spin polarization, measurable with, say, Knight shift
techniques, will increase from $P=0$ to $P=1$ according to the function
$P(\nu)=\frac{2}{2-\nu}-5$. 

  The unpolarized quasielectrons can manifest themselves in the I-V
characteristics for tunneling between two $\nu=1/3$ ferromagnets at
$B<20T$. In addition to the "polarized" characteristics, observable for
$B>20T$, a peak should appear at $V\approx 0.7 \frac{e}{\kappa l}$. This
peak is a signature of tunneling by the spin-up electrons.

\acknowledgements

J.D. was supported by UK PPARC and M.H. by Deutscher Akademischer
Austauschdienst.

\begin{table}
\caption{ Energy difference between the lowest states of $N$ electrons
          on a sphere with spin $S=N/2$ and $N/2-1$ at filling fraction 
         ``$1/3+1\;\text{quasiparticle}$''. The energy is in units of 
          $e^2/\varepsilon l'$ with $l'$ including a finite size 
          correction [9]
          $l'=l\protect\sqrt{\protect\frac{m \protect\nu}{N}}$,
          where $m$ is the number of flux quanta. }
 \label{t1}
 \begin{tabular}{ccc}
   N     & $E_{S=N/2} - E_{S=N/2-1}$ & $E_{S=N/2-1} - E_{S=N/2-2}$ \\ \hline
   5     &     0.04033               &    0.00610                \\
   6     &     0.03746               &    0.00624                \\
   7     &     0.03650               &    0.00607                      \\
   8     &     0.03547               &    0.00597                 \\
   9     &     0.03460               &                            \\
$\infty$ &     0.0278(5)             &    0.0052(1)
 \end{tabular}
 \end{table}

\begin{table}
\caption{ Magnetic fields at which various Coulomb energy eigenstates
          become stable for a model sample with $N=8$ electrons in spherical 
          geometry. }
 \label{t2}
 \begin{tabular}{ccccc}
         & $\Phi=1$ & $\Phi=2$ & $\Phi=3$ & $\Phi=4(\nu=2/5)$ \\ \hline
  $S=3$  &     42T  &   --     &    --    &   --              \\
  $S=2$  &     1.2T &   34.4T  &    --    &   --              \\
  $S=1$  &     --   &   0.4T   &    25.5T &   --              \\
  $S=0$  &     --   &   --     &    0.01T &   17.6T           \\ 
 \end{tabular}
 \end{table}

\end{document}